\documentclass[prd,preprintnumbers,superscriptaddress,showpacs,showkeys,nofootinbib]{revtex4-1}
\usepackage{amsmath,amsfonts,amssymb,amsthm,amstext,amscd,eucal,srcltx,mathrsfs}
\usepackage{epsfig,graphicx,bm}
\usepackage{epstopdf, epsf}
\usepackage{hyperref}
\usepackage{color}
\newcommand{\be}{\begin{equation}}
\newcommand{\ee}{\end{equation}}

\begin{document}
\title{Do Solar System experiments constrain scalar-tensor gravity?}
\author{Valerio Faraoni}
\email{vfaraoni@ubishops.ca}
\affiliation{Department of Physics \& Astronomy,
Bishop's University, 
2600 College Street, J1M 1Z7 Sherbrooke, Qu\'{e}bec, Canada}
\author{Jeremy C\^ot\'e}
\email{jcote16@ubishops.ca}
\affiliation{Department of Physics \& Astronomy,
Bishop's University, 
2600 College Street, J1M 1Z7 Sherbrooke, Qu\'{e}bec, Canada}
\affiliation{Perimeter Institute for Theoretical Physics, 31 Caroline 
Street North,Waterloo, ON N2L 2Y5, Canada}
\author{Andrea Giusti}
\email{agiusti@bo.infn.it}
\affiliation{Department of Physics \& Astronomy,
Bishop's University, 
2600 College Street, J1M 1Z7 Sherbrooke, Qu\'{e}bec, Canada}
\begin{abstract}

It is now established that, contrary to common belief, (electro-)vacuum 
Brans-Dicke gravity does not reduce to general relativity (GR) for 
large values of the Brans-Dicke coupling $\omega$. Since the essence of 
experimental tests of scalar-tensor gravity consists of providing lower 
bounds on $\omega$, in light of the misguided assumption of the 
equivalence between the limit $\omega \to \infty$ and the GR limit of 
Brans-Dicke gravity, the parametrized post-Newtonian (PPN) formalism on 
which these tests are based could be in jeopardy. We show that, in the 
linearized approximation used by the PPN formalism, the anomaly in the 
limit to general relativity disappears.  However, it survives to second 
(and higher) order and in strong gravity. In other words, while the 
weak gravity regime cannot tell apart GR and $\omega \to \infty$ 
Brans-Dicke gravity, when higher order terms in the PPN analysis of 
Brans-Dicke gravity are included, the latter never reduces to the one of 
GR in this limit. This 
fact is relevant for experiments aiming to test second order light 
deflection and Shapiro time delay.
 
\end{abstract}
\maketitle
\section{Introduction} 
\label{sec:1}
\setcounter{equation}{0}

Deviations from Einstein's theory of gravity, General Relativity (GR), 
appear in virtually all attempts to 
introduce quantum corrections to gravity 
\cite{corrections1}-\cite{corrections8} (for recent overviews of GR and 
the challenges it faces, see, {\em e.g.}, \cite{a,b,c}). 
In addition 
to these deviations (in the form of extra fields, higher order terms in  
the  
field equations, and non-minimal couplings to the curvature), compelling  
motivation to investigate alternatives to GR comes from the 1998 
discovery that the current expansion of the universe is  
accelerated. Within 
the standard $\Lambda$-Cold~Dark~Matter ($\Lambda$CDM) model of cosmology  
based on GR, one needs to introduce a completely {\em ad hoc} dark 
energy with a very exotic equation of state to explain the cosmic 
acceleration \cite{AmendolaTsujikawabook}. A popular alternative to dark 
energy consists of modifying 
gravity at large scales. Many modifications of GR have been proposed, the 
most studied being $f(R)$ gravity \cite{f(R)1, f(R)2}. This is a class of 
theories in which 
the Einstein-Hilbert Lagrangian density $R$ (the Ricci scalar of 
spacetime) is 
promoted to a non-linear function $f(R)$. It turns out 
\cite{reviews1, reviews2, reviews3} that 
this class of theories reduces to a Brans-Dicke theory with Brans-Dicke 
scalar  $\phi=f'(R)$, vanishing Brans-Dicke 
coupling parameter $\omega$, and the complicated potential 
$V(\phi)=Rf'(R)-f(R)\big|_{R=R(\phi)}$ (see Refs.~\cite{reviews1, 
reviews2, reviews3} for reviews and \cite{HarkoLobo} for extensions 
of $f(R)$ gravity). 

Brans-Dicke theory, originally introduced in Ref.~\cite{BD1, BD2, BD3} to 
account 
for Mach's principle, has been 
generalized to the wider class of scalar-tensor theories 
\cite{ST1, ST2, ST3} described by the 
action (we follow the 
notation of Ref.~\cite{Waldbook} and use units in which Newton's 
constant $G$ and the speed of light $c$ are unity) 
\be
S_{\rm ST} = \frac{1}{16\pi} \int d^4 x \sqrt{-g} \left[ \phi R 
-\frac{\omega(\phi )}{\phi} 
\, \nabla^c\phi \nabla_c\phi -V(\phi) \right]  +S^{\rm (m)} \,, 
\label{STaction}
\ee
where the Brans-Dicke scalar $\phi$ corresponds approximately to the 
inverse of the gravitational coupling strength $G_{\rm eff}$, $\omega$ is 
the 
Brans-Dicke coupling, and $V(\phi) $ is a potential for $\phi$, 
which gives a range to this field. $S^{\rm (m)}$ is the matter action. 
Besides 
containing the 
cosmologically motivated class of $f(R)$ theories, scalar-tensor 
gravity, which adds only a (massive) scalar degree of freedom $\phi \simeq 
G_{\rm eff}^{-1}$ to the massless spin two graviton of GR,   
constitutes a minimal modification of GR and is the prototype of the 
alternative theory of gravity \cite{Will1, Will2, Will3}. The field 
equations are 
\cite{BD3, ST1, ST2, ST3} 
\begin{eqnarray}
R_{ab} - \frac{1}{2}\, g_{ab} R &=& \frac{8\pi}{\phi} \,  T_{ab}^{\rm (m)} 
+ \frac{\omega}{\phi^2} \left( \nabla_a \phi 
\nabla_b \phi -\frac{1}{2} \, g_{ab} 
\nabla_c \phi \nabla^c \phi \right) \nonumber\\
&&\nonumber\\
&\, &  +\frac{1}{\phi} \left( \nabla_a \nabla_b \phi 
- g_{ab} \Box \phi \right) 
-\frac{V}{2\phi}\, 
g_{ab} \,,\label{BDfe1}\\
&& \nonumber\\
\Box \phi = \frac{1}{2\omega+3} & & 
\left( \frac{8\pi T^{\rm (m)} }{\phi}   + \phi \, \frac{d V}{d\phi} 
-2V -\frac{d\omega}{d\phi} \nabla^c \phi \nabla_c \phi \right)  \,,
\label{BDfe2}
\end{eqnarray}
where $R_{ab}$ is the Ricci tensor and $\nabla_a $ is the covariant 
derivative of the spacetime metric $g_{ab}$, while $ T^{\rm (m)} \equiv 
g^{cd}T_{cd}^{\rm (m)} $ is the trace of 
the matter energy-momentum tensor $T_{ab}^{\rm (m)}=-\frac{2}{\sqrt{-g}} 
\, 
\frac{\delta S^{\rm (m)}}{\delta g^{ab}} $.

Scalar-tensor gravity in the Jordan frame $\left( g_{ab}, \phi \right)$, 
can be reformulated in the Einstein conformal frame 
$( \tilde{g}_{ab}, \tilde{\phi} )$ as follows \cite{BD3}. 
Perform 
the conformal transformation of the  metric 
tensor 
\be \label{metrictransformation} g_{ab} 
\rightarrow \tilde{g}_{ab} \equiv \phi \, g_{ab} \,, 
\ee 
and the scalar field redefinition 
\be 
d\tilde{\phi} = \sqrt{ 
\frac{|2\omega+3|}{16\pi}} \, \frac{d\phi}{\phi} \,. 
\ee 
Since we restrict ourselves to Brans-Dicke theory with constant $\omega$, 
we have the non-linear scalar field redefinition 
\be 
\phi \rightarrow 
\tilde{\phi}=\sqrt{\frac{|2\omega+3|}{16 \pi}} \, \ln \left( 
\frac{\phi}{\phi_0}\right) \,, \label{phitransform} 
\ee 
where $\phi_0$ is an integration constant and $\omega\neq -3/2$. Both 
$\phi$ and $g_{ab}$ depend on the parameter $\omega$, therefore 
the Einstein frame metric $\tilde{g}_{ab}$ in general depends on 
the parameter $\omega$ (the same is true, in general, for the 
Einstein frame scalar $\tilde{\phi}$).

Using the Einstein frame variables $( \tilde{g}_{ab}, \tilde{\phi})$, the  
Brans-Dicke action~(\ref{STaction}) (with $\omega=$~const.)  
is rewritten as
\be
S_{\rm BD} = \int d^4x\sqrt{-\tilde{g}}\left[
\frac{\tilde{R }}{16 \pi}-\frac{1}{2} \, 
\tilde{g}^{ab}\nabla_a\tilde{\phi} \nabla_b\tilde{\phi}  
-U(\tilde{\phi})  + \frac{{\cal  
L}^{\rm (m)}}{\phi^2(\tilde{\phi}) }  \right]  \,, 
\label{BDactionEframe} 
\ee
where
\be
U (\tilde{\phi}) = 
\left. \frac{V(\phi)}{16\pi \phi^2}\right|_{\phi=\phi( \tilde{\phi}) }  \,,  \label{ppotential} 
\ee
where we denote Einstein frame quantities with a 
tilde. Formally, this is the 
Einstein-Hilbert action of GR with a matter scalar field 
with canonical kinetic energy density, but  
this scalar $\tilde{\phi}$ now couples non-minimally to matter. In the 
Einstein frame, the Brans-Dicke field equations become
\begin{eqnarray}
\tilde{R}_{ab}-\frac{1}{2} \, \tilde{g}_{ab} \tilde{R}  &=& 
8\pi \left(  \mbox{e}^{- \sqrt{\frac{64\pi}{|2\omega+3|}} 
\, \tilde{\phi} } \, T_{ab}^{\rm (m)} + 
\tilde{\nabla}_a \tilde{\phi} \tilde{\nabla}_b
\tilde{\phi} 
-\frac{1}{2} \, \tilde{g}_{ab}\, 
\tilde{g}^{cd} \tilde{\nabla}_c \tilde{\phi} \tilde{\nabla}_d \tilde{\phi} 
\right. \nonumber\\
&&\nonumber\\
&\, & \left.  - U(\tilde{\phi})  \, \tilde{g}_{ab} \right)   
\,, \label{Eframefe}
\end{eqnarray}
\be
 \tilde{g}^{ab} \tilde{ \nabla}_a  \tilde{ \nabla}_b
\tilde{\phi} -\frac{dU}{d\tilde{\phi}} 
+8\, \sqrt{ \frac{\pi}{|2\omega+3|}} \,  \mbox{e}^{ 
-\sqrt{\frac{64\pi}{|2\omega+3|}}\, \tilde{\phi} } \, {\cal 
L}^{\rm (m)} = 0 \,. 
\label{EframeKG}
\ee
From now on we restrict to vacuum Brans-Dicke 
theory and set $T_{ab}^{\rm (m)}=0$.  The 
explicit coupling  between Einstein frame scalar $\tilde{\phi}$ and 
matter then disappears and the Einstein frame 
action~(\ref{BDactionEframe})  
is formally the Einstein-Hilbert action and the 
Einstein frame pair $( \tilde{g}_{ab}, \tilde{\phi} )$ 
is formally a scalar field  solution of the Einstein equations even though   
it has been generated by the original  Jordan 
frame spacetime  $\left( g_{ab}, \phi \right)$.

\section{$\omega\rightarrow \infty$ vs. GR limit}
\label{sec:2}
\setcounter{equation}{0}

In practice, Brans-Dicke theory with $\omega=$~const.  
is used to approximate all scalar-tensor theories in experimental tests of 
gravity in the weak field regime \cite{Will1, Will2, Will3} (this 
situation can be 
different in strong gravity when scalarization is involved, but we are 
not concerned with this type of situation here). 
It is clear that Brans-Dicke gravity reduces to GR if $\phi$ becomes 
constant. Precisely, the {\em GR limit of Brans-Dicke gravity} is 
understood as
the limit in which Brans-Dicke gravity coupled to matter reduces to GR 
sourced by 
the same type of matter. 

The belief that $\phi$ does so in the limit  $\omega\rightarrow \infty $ 
is standard textbook material ({\em e.g.}, \cite{Weinberg}).  However, 
the asymptotics 
of $\phi$ in this limit are important. While in most cases these asymptotics are 
$\phi=\phi_{\infty}+\mathcal{O}( 1/\omega)$, where 
$\phi_{\infty}$ is a 
constant \cite{Weinberg}, many analytic solutions of the Brans-Dicke 
field equations have been discovered 
over the years 
for which $\phi=\phi_{\infty}+\mathcal{O}( 1/\sqrt{|\omega|})$, which do 
not go 
over to the corresponding GR solutions with the same form 
of matter \cite{failure1, failure2, failure3, failure4, failure5, 
failure6, failure7, BhadraNandi}.\footnote{Similar anomalies are 
occasionally reported for instances of Brans-Dicke solutions with 
non-conformal matter \cite{Chauvineau1, Chauvineau2, newpapers1, 
newpapers2, Brando}.} 
Far from being limited to 
a few maverick solutions, this problem has later been shown to affect {\em 
the entire electrovacuum} ({\em i.e.}, $T^{\rm (m)}=0$) {\em theory}  
\cite{BanerjeeSen97} and a formal explanation has been given for this 
``anomalous'' behaviour \cite{BanerjeeSen97, myBDlimit1, myBDlimit2}.

Deviations from GR are well constrained experimentally in the Solar 
System, where gravity is weak, and to some extent also outside of it 
\cite{Will1, Will2, Will3, tests1, tests2}. Assuming the Brans-Dicke field 
to be 
long-ranged, the 
best limits on scalar-tensor gravity arise from the Cassini probe and are 
$|\omega|> 40000$ \cite{BertottiIessTortora}. In general, experiments 
provide a lower bound on $|\omega|$, constraining this parameter to be 
large (unless $\phi$ becomes so massive and short-ranged to escape this 
limit, as in viable $f(R)$ models \cite{reviews1, reviews2, reviews3}).

The Solar System experiments probe gravity in vacuo, the situation in 
which the  $\omega\rightarrow \infty $ limit is anomalous. Therefore, how 
can experiments constraining the deviations from GR in the 
field of the Sun and forcing 
$|\omega|$ to be large, apply to a theory that does not reduce 
to GR in this limit? Can the Parametrized 
Post-Newtonian (PPN) approximation, which constitutes the basis for 
analyzing these experiments \cite{Will1, Will2, Will3, Iorio1, Iorio2, 
Iorio3},
still discriminate between GR and Brans-Dicke gravity in the large 
$\omega$ regime?

This question is crucially important for experimental tests of 
scalar-tensor gravity, but it has not been posed in the literature thus 
far. Here we provide an answer: the exact (strong gravity) electrovacuum 
theory definitely does not reduce to GR as $\omega \rightarrow \infty$. In 
this limit, a (canonical, minimally coupled) scalar field survives in the 
limit of the field equations and acts as a matter source  
\cite{usPRD,mycosmobook}. 
However, the PPN analysis is 
limited to the 
weak field expansion of these field equations and, in this regime, the 
offending terms disappear from these equations, in which the dominant 
terms introduced by the scalar degree of freedom $\phi$ conform, instead, 
to the usual PPN 
analysis. This simplification occurs only to first order in the deviations 
of the metric and Brans-Dicke scalar from the Minkowski background, 
and are 
bound to reappear to second order and, of course, in any exact 
(strong gravity) electrovacuum solution of the theory. This fact is of 
interest for future experiments 
testing light deflection and Shapiro time delay to second order 
\cite{2ndorderlight1}-\cite{2ndorderlight4}.

Now to the technical details. It is clear that, if the Brans-Dicke 
scalar becomes constant, Brans-Dicke gravity reduces to GR  and therefore 
one should recover $\phi 
\rightarrow $~const. as $\omega\rightarrow \infty$.  The rate at which  
$\phi$ approaches a constant is important. The  
gradient $\nabla\phi$ decays as  $|\omega |$ becomes larger,  
\be
\nabla_a \phi =  \phi_0 \sqrt{\frac{16\pi}{|2\omega+3|}} 
\, \, \exp \left(  \sqrt{\frac{16\pi}{|2\omega+3|}}  \,\tilde{\phi} 
\right)  \, \tilde{\nabla}_a \tilde{\phi} \,. \label{stucz}
\ee
Consider the Jordan 
frame field equations and, in particular,   
the term which appears in their right hand side
\be
A_{ab} \equiv \frac{\omega}{\phi^2} \left( \nabla_a  
\phi \nabla_b \phi -\frac{1}{2} \, g_{ab} \, \nabla^c  
\phi \nabla_c \phi \right) \,.
\ee
In the literature, the failure of Jordan frame Brans-Dicke theory to 
reproduce the expected GR limit (which corresponds to $\phi=$~const. and 
to the vanishing of the right hand side of the vacuum field equations) has 
been reognized to follow from 
the fact that, when the asymptotics is given by 
$\phi=\phi_{\infty}+\mathcal{O}( 1/\sqrt{|\omega|})$, the tensor 
$A_{ab}$ does not vanish in the $\omega \rightarrow \infty 
$ limit but remains of order unity \cite{failure1, failure2, failure3, failure4,
failure5, failure6, failure7, BanerjeeSen97}. It is easy to see that, when 
Eq.~(\ref{stucz}) is true, 
the tensor $A_{ab}$   reads 
\begin{eqnarray}
A_{ab} &=& \frac{\omega}{\phi^2} \, \phi_0^2 \,  
\mbox{e}^{ 2\sqrt{\frac{16\pi}{|2\omega+3|}} 
\,\tilde{\phi} } 
\frac{16\pi}{|2\omega+3|}  \left( \nabla_a
\tilde{\phi} \nabla_b \tilde{\phi} -\frac{1}{2} \, 
g_{ab} 
g^{cd} \, \nabla_c \tilde{\phi} \nabla_d 
\tilde{\phi} \right) \\  
&&\nonumber\\  
& = &  16 \pi \, \mbox{sign} (\omega) \left| \frac{\omega}{2\omega+3} 
\right| 
\left( \frac{\phi_0}{\phi} \right)^2 \, \mbox{e}^{ 2\sqrt{ 
\frac{16\pi}{ |2\omega+3|} } } \Big( \nabla_a\tilde{\phi} \nabla_b 
\tilde{\phi} 
- \frac{1}{2} \, \frac{\tilde{g}_{ab} }{\phi} \, 
\phi \, \tilde{g}^{cd}  \nabla_c \tilde{\phi} \nabla_d \tilde{\phi} 
\Big)   
\end{eqnarray}
for all values of the parameter $\omega$. Now, in the limit  
$\omega\rightarrow \infty$  in which 
$\phi\rightarrow \phi_0$, one obtains
\be
A_{ab} \rightarrow  A_{ab}^{(\infty)} = 
8\pi \, \mbox{sign}(\omega) \left( \tilde{\nabla}_a
\tilde{\phi} \tilde{\nabla}_b \tilde{\phi} -\frac{1}{2} \, 
\tilde{g}_{ab} \, \tilde{g}^{cd} 
\, \tilde{\nabla}_c \tilde{\phi} \tilde{\nabla}_d \tilde{\phi} \right) \,.
\label{stucz2}
\ee 
The Einstein frame metric $\tilde{g}_{ab}^{(\infty)}$  solves the Einstein 
equations with the scalar field $\tilde{\phi}$ as the only matter 
source. This field has canonical stress-energy tensor 
$A_{ab}^{(\infty)}$, which is obtained as the limit of the {\em Jordan 
frame} stress-energy tensor, as
\be
\left.
8\pi \tilde{T}_{ab}[\tilde{\phi}]  
\right|_{\mbox{{\scriptsize Einstein~frame}}} = 
\left.
A_{ab}^{(\infty)} \right|_{\mbox{{\scriptsize Jordan~frame~limit}}}   
\ee
For $\omega>0$, this Einstein frame scalar  couples  minimally to the 
curvature and has canonical kinetic energy density. One obtains the same 
metric tensor by considering two candidates for a GR limit of 
Brans-Dicke 
theory: the $\omega\rightarrow \infty$ limit of the Einstein 
frame metric  and the 
$\omega\rightarrow \infty$ limit of the Jordan frame metric, which 
coincide apart from an irrelevant positive multiplicative constant 
$\phi_{\infty}$. However, this metric  
$\tilde{g}_{ab}^{(\infty)}=g_{ab}^{(\infty)}$  obtained with these 
two  different methods is not a solution of the vacuum 
Einstein equations (this would require instead $A_{ab}^{(\infty)}$ to 
vanish identically). Instead,  
$\tilde{g}_{ab}^{(\infty)}=g_{ab}^{(\infty)}$  solves the coupled 
Einstein-Klein-Gordon equations and, therefore, vacuum Brans-Dicke theory 
does not reproduce vacuum GR in the limit, as it should be for a correct 
``limit to GR''.

\section{PPN analysis and Brans-Dicke anomaly}

It is well known that the only stationary, spherically symmetric, 
asymptotically flat black hole solution of Brans-Dicke gravity with 
$V(\phi) = 0$ is the Schwarzschild metric \cite{nohair1}-\cite{nohair4}. 
If one assumes 
the absence of an event horizon, however, the most general static, 
spherically symmetric,  asymptotically flat solution of the vacuum 
Brans-Dicke field equations with 
vanishing potential is parametrized by three continuous real parameters 
$(\alpha _0 , \beta_0 , \gamma)$ (see, {\em e.g.}, \cite{JB, 
FaraoniCote2018} and references therein). In detail, for $\gamma \neq 0$ 
the general solution reads
\be \label{gno0}
 d s ^2 _{\gamma \neq 0} =
- \mbox{e}^{(\alpha_0 + \beta_0)/r} \, d t^2  
+ \mbox{e}^{(\beta_0 - \alpha_0 )/r} \, 
\left(\frac{\gamma/r}{\sinh(\gamma/r)} 
\right)^4 \, d r^2 + \mbox{e}^{(\beta_0 - \alpha_0 )/r} \, 
\left(\frac{\gamma/r}{\sinh(\gamma/r)} \right)^2 \, r^2 d \Omega_{(2)}^2  \, ,
\ee

\be
\phi (r) = \phi _0 \, \mbox{e}^{- \beta_0 /r} \, \, , \quad \quad\quad 
\beta _0 = \frac{\sigma}{\sqrt{| 2 \, \omega + 3|} }\,,
\ee
where $d \Omega_{(2)}^2 \equiv d \theta ^2 + \sin^2 \theta \, d 
\varphi^2$, $\sigma$ denotes a scalar charge, and $
4 \, \gamma^2 = \alpha_0^2 + 2 \, \sigma^2$.\footnote{The last 
condition only holds for $\gamma > 0$ and there is no loss of generality 
in choosing $\gamma >0$ when $\gamma \neq 0$.} If instead $\gamma = 0$, 
the  solution is given by the Brans class~IV spacetime 
\cite{Brans, FaraoniCote2018})
\be \label{g0}
d s ^2_{(0)} =
- \mbox{e}^{(\alpha_0 + \beta_0)/r} \, d t^2  
+ \mbox{e} ^{(\beta_0 - \alpha_0 )/r} \, \left(d r^2 + r^2 d 
\Omega_{(2)}^2  \right) 
\, ,
\ee

\be
\phi (r) = \phi _0 \, \mbox{e}^{- \beta_0 /r} \, . 
\ee

It is easy to see that, for this class of solutions, the scalar field 
approaches a constant value $\phi _\infty = \phi _0$ in the limit $\omega 
\to \infty$ as
\be \label{asymphi}
\phi (r) \sim \phi _0 - \frac{\phi _0 \, \sigma}{\sqrt{2 |\omega|} \, r} + 
\mathcal{O}\left(\frac{1}{|\omega|}\right) \, ,
\ee
which is indeed the typical behavior for which the anomaly comes up. 

Now, for $\omega \to \infty$, Eq.~(\ref{gno0}) reduces to the 
Fisher-Janis-Newman-Winicour-Buchdahl-Wyman metric in the limit, which is 
known to be the general static, spherical, asymptotically flat solution of 
the Einstein field equations sourced by a free scalar field $\Phi (r) = - 
\sigma / (4 \sqrt{\pi} r)$, featuring a naked singularity at its centre.  
Alternatively, when $\gamma=0$ Eq.~(\ref{g0}) approaches the Yilmaz 
geometry as $\omega 
\to \infty$, which is again a solution of the Einstein field equations 
with a scalar field $\Phi (r) \propto 1 / r$. In other words, as $\omega 
\to \infty$ the family of general solutions of {\em vacuum} Brans-Dicke 
gravity 
discussed above {\em do not reduce} to a general solution of the {\em 
vacuum} 
Einstein field equations (namely the Minkowski space, since we have 
assumed 
the absence of event horizons). Instead, this family of 
solutions approaches two families of spacetimes corresponding to {\em 
non-vacuum} solutions of the Einstein field equations, thus breaking the 
equivalence between the GR limit of Brans-Dicke gravity and the limit 
$\omega \to 
\infty$.

It is then interesting to see how the PPN analysis of scalar-tensor 
theories is affected by this anomalous behavior. Using the wisdom coming 
from the general static, spherical, asymptotically flat non-black-hole 
class of solutions of {\em vacuum} Brans-Dicke gravity one can show that, 
whilst 
the equivalence of the two limits is not affected at the first 
post-Newtonian order, an effective scalar field stress-energy tensor 
survives at the next-to-leading order in the $\omega\rightarrow \infty$ 
limit. This, in turn, {\em prevents the full Brans-Dicke theory from 
reducing to GR in this limit}.

In the weak field limit the metric and scalar field are expanded as 
\begin{eqnarray}
g_{\mu\nu}&=&\eta_{\mu\nu} + h_{\mu\nu} 
\,,\\
&&\nonumber\\
g^{\mu\nu}&=&\eta^{\mu\nu} - h^{\mu\nu} + \frac{1}{2} \, h^{\mu \alpha} 
h_{\alpha}^{\,\, \nu} + \mathcal{O} (h^3) 
\,,\\
&&\nonumber\\
\phi &=&\phi_0 + \varphi + \frac{\varphi ^2}{2} + \mathcal{O} (\varphi^3) 
\,,
\end{eqnarray}
where $\eta_{\mu\nu}$ is the Minkowski metric and $\phi_0$ is a constant, 
while $h_{\mu\nu}$ and $\varphi$ are small perturbations. From 
Eqs.~(\ref{gno0}) and~(\ref{g0}) one infers that
$$
h_{\mu \nu} \sim \frac{\alpha _0 \pm \beta _0}{r}  
$$
in the weak field limit. Besides, since $\beta_0 \sim \sigma / \sqrt{2 \, 
|\omega|}$ for large $\omega$, one can conclude that
\be \label{asymph}
 h_{\mu \nu} \sim \frac{\alpha _0}{r} \pm \frac{\sigma}{\sqrt{2  
|\omega|} \, r}  
\qquad \mbox{as} \,\,\, \omega \to \infty \, , 
\ee
which further implies that the anomaly does not show up in the weak field 
expansion of the left hand side of Eq.~(\ref{BDfe1}) since this  
contains only positive powers of $h_{\mu \nu}$ and its derivatives. 
However, 
the right hand side of Eq.~(\ref{BDfe1}) for $V (\phi) = 0$ and in {\em 
vacuo} has a peculiar behavior. Indeed, expanding this term up to  
third order one finds
\begin{eqnarray} 
\frac{\omega}{\phi^2} \Big( \partial_\mu \phi  \partial_\nu 
\phi &-&  \frac{1}{2} \, g_{\mu \nu} \,  \partial_\alpha \phi 
\partial^\alpha \phi \Big)
+\frac{\nabla_\mu \partial_\nu \phi}{\phi}\nonumber\\
&&\nonumber\\
& &=  \frac{\omega}{\phi _0 ^2} \Big( \partial_\mu \varphi  
\partial_\nu \varphi -
\frac{1}{2} \, \eta_{\mu \nu} \,  \partial_\alpha \varphi \partial^\alpha 
\varphi \Big) +\frac{1}{\phi_0} \Big[ \partial_\mu \partial _\nu \varphi + 
\partial_\mu \partial _\nu (\varphi^2 / 2) \nonumber\\
&&\nonumber\\
& & + \big(\partial_\mu h_{\nu \alpha} + \partial_\nu h_{\mu \alpha} - 
\partial_\alpha h_{\mu \nu} \big) 
\partial ^\alpha \varphi \Big]  
+ \mathcal{O} \Big( \varphi ^3  \, , \,  h 
\partial \varphi^2 \, , \,  h^2 \partial \varphi \Big) \,.\nonumber\\
&& \label{exp}
\end{eqnarray}
Now, assuming that we work within the scenario that leads to the exact 
solutions discussed above and using the asymptotics~(\ref{asymphi}) 
and~(\ref{asymph}) it is easy to see that, while at the first post-Newtonian 
order the scalar field contribution disappears, to second order in 
the PPN expansion the first term in the left hand side of Eq.~(\ref{exp}) 
is $\mathcal{O} (\omega ^0)$ and survives the limit $\omega \to \infty$, 
breaking the equivalence between these two limits.

	Let us now make more quantitative predictions using the line element \eqref{gno0} as an example. First, in \eqref{gno0} one identifies
the areal radius
\be
R (r) = 
\mbox{e}^{(\beta_0 - \alpha_0 )/2 r} \, 
\frac{\gamma/r}{\sinh(\gamma/r)} \, r \, .
\ee
Expanding for large $r$ one then finds
\be
R = r + \frac{\beta_0 - \alpha_0}{2} + 
\frac{3 (\beta_0 - \alpha_0 )^2 - 4 \gamma^2}{24 \, r}
+ \mathcal{O} \left( \frac{1}{r^2} \right) \, ,
\ee	
that implies
\be
d r^2 \simeq \left( 1 + \frac{3 (\beta_0 - \alpha_0 )^2 - 4 \gamma^2}{12 \, r^2}  \right) \, d R^2 \, .
\ee
Hence one can implicitly recast the line element \eqref{gno0} in terms of the areal radius as
\be
d s ^2 _{\gamma \neq 0} =
g_{tt} \, d t^2  
+ g_{RR} \, d R^2 + R^2 \, r^2 d \Omega_{(2)}^2  \, ,
\ee
with
\be
g_{tt} = - \mbox{e}^{(\alpha_0 + \beta_0)/r}  = - \left( 1 + \frac{\alpha_0 + \beta_0}{r} + \frac{(\alpha_0 + \beta_0)^2}{2 \, r^2} \right) +
\mathcal{O} \left( \frac{1}{r^3} \right) \, ,
\ee
\begin{eqnarray}
\nonumber g_{RR} &=& \mbox{e}^{(\beta_0 - \alpha_0 )/r} \, 
\left(\frac{\gamma/r}{\sinh(\gamma/r)} 
\right)^4 \left[ 1 + \frac{3 (\beta_0 - \alpha_0 )^2 - 4 \gamma^2}{12 \, r^2} + 
\mathcal{O} \left( \frac{1}{r^3} \right) \right]\\
&=&
1 + \frac{\beta_0 - \alpha_0}{r} + \frac{3 (\beta_0 - \alpha_0 )^2 - 4 \gamma^2}{4 \, r^2} +
\mathcal{O} \left( \frac{1}{r^3} \right)
\end{eqnarray}
and $r=r(R)$\footnote{Note that for large values of $r$ one has $R \simeq r$.}.

	Performing the usual PPN identifications
	\be
	g_{tt} = - (1 + 2 \, \Psi(r)) \quad \mbox{and} \quad
	g_{RR} = 1 + 2 \, \Phi (r) \, ,
	\ee
the post-Newtonian parameter $\gamma _{\rm PPN}$ (not to be confused with $\gamma$) reads
\be
\gamma _{\rm PPN} = - \frac{\Psi(r)}{\Phi(r)} = 
\frac{\alpha_0 + \beta_0}{\alpha_0 - \beta_0} 
\left( 1 + \frac{5\alpha_0 ^2 - 6 \alpha_0 \beta_0 + \beta_0 ^2 - 4 \gamma^2}{4 \, (\alpha_0 - \beta_0) \, r} \right) + \mathcal{O} \left( \frac{1}{r^2} \right) \, .
\ee
Taking the limit $\omega \to \infty$, {\em i.e.} $\beta _0 \to 0$, one finds
\be
\lim _{\omega \to \infty} (\gamma _{\rm PPN} - 1) = 
\frac{5\alpha_0 ^2 - 4 \gamma^2}{4 \, \alpha_0 \, r} + \mathcal{O} \left( \frac{1}{r^2} \right) \, ,
\ee	
which is always non-vanishing for a scalar charge $\sigma \neq 0$.

\section{Conclusions}
\label{sec:3}

As a conclusion, the PPN analysis narrowly escapes the problem of the GR 
limit 
arising in the full theory. It is clear, however, that  this problem will 
reappear as soon as second and higher order terms are included in the 
weak field expansion and, of course, in the full strong gravity 
regime. To second order, the PPN analysis of scalar-tensor gravity is 
in jeopardy. The divergence between PPN predictions and the  
$\omega\rightarrow \infty $ limit of Brans-Dicke theory will then be relevant.
In particular, this divergence will become important in the experimental determination of light deflection by 
the gravitational field of the Sun to second order in the PPN expansion 
\cite{2ndorderlight1}-\cite{2ndorderlight4}. These deviations could be 
obtained, in principle, 
with high 
precision astrometry, in testing strong gravity effects with the Event 
Horizon Telescope \cite{ETH1, ETH2} and, potentially, in tests based on 
gravitational waves \cite{gwtests1}-\cite{gwtests7}. 
Such strong gravity effects, which look more promising for detecting
scalar-tensor gravity effects or further constraining the theory, will be explored in future work.

\section*{Acknowledgments} This work is supported, in part, by the Natural 
Science and Engineering Research Council of Canada (Grant No. 2016-03803 
to V.F.) and by Bishop's University. This research was supported, in part, 
by  the Perimeter Institute for Theoretical Physics. Research at Perimeter 
Institute is supported by the Government of Canada through Industry Canada 
and by the Province of Ontario through the Ministry of Economic 
Development and Innovation.


\begin{thebibliography}{99}

\bibitem{corrections1} R. Utiyama and B.S. DeWitt, {\em J. Math. Phys.} 
{\bf 3}, 608 (1962).

\bibitem{corrections2} K.S. Stelle, {\em Gen. Relat. Gravit.} {\bf 9}, 353 
(1978).

\bibitem{corrections3}C.G. Callan, D. Friedan, E.J. Martinez, and M.J. 
Perry, {\em Nucl. Phys. B} {\bf 262}, 593 (1985).

\bibitem{corrections4}E.S. Fradkin and A.A. Tseytlin, {\em Nucl. Phys. B} 
{\bf 261}, 1 (1985).

\bibitem{corrections5} A. Giusti, {\em Int. J. Geom. Meth. Mod. Phys.} 
{\bf 16}, 1930001 (2019).

\bibitem{corrections6} R. Casadio, A. Giugno, and 
A. Giusti, {\em Phys.\ Rev.  D} {\bf 97}, 024041 (2018).

\bibitem{corrections7}I.L. Buchbinder, S.D. Odintsov, and I.L. Shapiro, 
{\em Effective Actions in Quantum Gravity} (IOP, Bristol, 1992).

\bibitem{corrections8} G.A. Vilkovisky, {\em Classical Quantum Gravity} 
{\bf 9}, 895 (1992).

\bibitem{AmendolaTsujikawabook} L. Amendola and S. Tsujikawa, {\em Dark 
Energy, Theory and Observations} (Cambridge University Press, Cambridge, 
2010).

\bibitem{f(R)1} S. Capozziello, S. Carloni, and A. Troisi, {\em Recent 
Res. Dev. Astron. Astrophys.} {\bf 1}, 625 (2003) 
[arXiv:astro-ph/0303041].

\bibitem{f(R)2} S.M. Carroll, V. Duvvuri, M. Trodden, and M.S. Turner, 
{\em Phys. Rev. D} {\bf 70}, 043528 (2004).

\bibitem{reviews1} T.P. Sotiriou and V. Faraoni, {\em Rev. Mod. Phys.} 
{\bf 82}, 451 (2010).

\bibitem{reviews2} A. De Felice and S. Tsujikawa, {\em Living Rev. 
Relativity} {\bf 13}, 3 (2010).

\bibitem{reviews3}S. Nojiri and S.D. Odintsov, {\em Phys. Rep.} {\bf 505}, 
59 (2011).

\bibitem{HarkoLobo} T. Harko and F.S.N. Lobo, {\em Extensions of 
$f(R)$  Gravity} (Cambridge University Press, Cambridge, 2018).

\bibitem{BD1} P. Jordan, {\em Naturwiss.} {\bf 26}, 417 
(1938).

\bibitem{BD2} P. Jordan, {\em Z. Phys.} {\bf 157}, 112 (1959).

\bibitem{BD3} C.H. Brans and R.H. Dicke, {\em Phys. Rev.} {\bf 124}, 925 
(1961).

\bibitem{ST1} P.G. Bergmann, {\em Int. J. Theor. Phys.} {\bf 1}, 25 
(1968).

\bibitem{ST2} R.V. Wagoner, {\em Phys. Rev. D} {\bf 1}, 3209 (1970).

\bibitem{ST3} K. Nordtvedt, {\em Astrophys. J.} {\bf 161}, 1059 (1970).

\bibitem{Waldbook} R.M. Wald, {\em General Relativity} (Chicago University 
Press, Chicago, 1984).

\bibitem{Will1} C.M. Will, {\em Theory and Experiment in Gravitational 
Physics} (Cambridge University Press, Cambridge, 1993).

\bibitem{Will2} C.M. Will, {\em Living Rev. Relativ.} {\bf 9}, 3 (2006).

\bibitem{Will3} E. Poisson and C.M. Will. {\em Gravity: Newtonian, 
Post-Newtonian, Relativistic} (Cambridge University Press, Cambridge, 
2014).

\bibitem{Iorio1} L. Iorio, N. Radicella, and M.L.  Ruggiero, {\em J. 
Cosmol. Astropart. Phys.} {\bf 08}, 021 (2015).
 
\bibitem{Iorio2} L. Iorio, {\em Int. J. Mod. Phys. D} {\bf 23}, 1450006 
(2014).
 
\bibitem{Iorio3} L. Iorio, {\em J. Cosmol. Astropart. Phys.} {\bf 07}, 001 
(2012).

\bibitem{Weinberg} S. Weinberg, {\em Gravitation and Cosmology} (Wiley, 
New York, 1972).

\bibitem{failure1} T. Matsuda, {\em Progr. Theor. Phys.} {\bf 47}, 738 
(1972).

\bibitem{failure2} C. Romero and A. Barros, {\em Astrophys. Sp. Sci.} {\bf 
192}, 263 (1992).

\bibitem{failure3} C. Romero and A. Barros, {\em Phys. Lett. A} {\bf 173}, 
243 (1993).

\bibitem{failure4} C. Romero and A. Barros, {\em Gen. Relativ. Gravit.} 
{\bf 25}, 491 (1993).

\bibitem{failure5}F.M. Paiva and C. Romero, {\em Gen. Relativ. Gravit.} 
{\bf 25}, 1305 (1993).

\bibitem{failure6} L.A. Anchordoqui, D.F. Torres, M.L. Trobo, and S.E. 
Perez-Bergliaffa, {\em Phys. Rev. D} {\bf 57}, 829 (1998).

\bibitem{failure7} P. Kirezli and \"{O}. Delice, {\em Phys. Rev. D} {\bf 
92}, 104045 (2015).

\bibitem{BhadraNandi} A. Bhadra and K.K. Nandi, {\em Phys. Rev. D} {\bf 
64}, 087501 (2001).

\bibitem{Chauvineau1} B. Chauvineau, {\em Classical Quantum Gravity} {\bf 
20}, 2617 (2003).

\bibitem{Chauvineau2} B. Chauvineau, {\em Gen. Relativ. Gravit.} {\bf 39}, 
297 (2007).

\bibitem{newpapers1} L. J\"arv, P. Kuusk, and M. Saal, {\em Phys. Rev. D} 
{\bf 76}, 103506 (2007).

\bibitem{newpapers2}  L. J\"arv, P. Kuusk, M. Saal, and O. Vilson, {\em 
Classical Quantum Gravity} {\bf 32}, 235013 (2015).

\bibitem{Brando} G. Brando, J.C. Fabris, F.T. Falciano, and O. 
Galkina, {\em  Int. J. Mod. Phys. D} {\bf 28}, 1950156 (2019).

\bibitem{BanerjeeSen97} N. Banerjee and S. Sen, {\em Phys. Rev. D} {\bf 
56}, 1334 (1997).

\bibitem{myBDlimit1} V. Faraoni, {\em Phys. Lett. A} {\bf 245}, 26 (1998). 

\bibitem{myBDlimit2} V. Faraoni, {\em Phys. Rev. D} {\bf 59}, 084021 
(1999).

\bibitem{tests1} E. Berti {\em et al.}, {\em Classical Quantum Gravity} 
{\bf 32},  243001 (2015).

\bibitem{tests2} T. Baker, D. Psaltis, and C. Skordis, {\em Astrophys. J.} 
{\bf 802}, 63 (2015).

\bibitem{BertottiIessTortora} B. Bertotti, L. Iess, and P. Tortora, {\em 
Nature} {\bf 425}, 374 (2003).

\bibitem{usPRD} V. Faraoni and J. C\^ot\'e, {\em Phys. Rev. D} {\bf 99}, 
064013 (2019).

\bibitem{2ndorderlight1} R. Epstein and I. I. Shapiro, {\em Phys. Rev. D} 
{\bf 22}, 2947 (1980).

\bibitem{2ndorderlight2} E. Fischbach and B.S. Freeman, {\em Phys. Rev. D} 
{\bf 22}, 2950 (1980).

\bibitem{2ndorderlight3} G.W. Richter and R.A. Matzner, {\em Phys. Rev. D} 
{\bf 26}, 1219 (1982).

\bibitem{2ndorderlight4} J. Bodenner and C.M. Will, {\em Am. J. Phys.} 
{\bf 71}, 8 (2003).


\bibitem{FaraoniCote2018} V. Faraoni and J. C\^ot\'e, {\em Phys. Rev. D} 
{\bf 98}, 084019 (2018).

\bibitem{Brans} C.H. Brans, {\em Phys. Rev.} {\bf 125}, 2194 (1962).

\bibitem{mycosmobook} V. Faraoni, {\em Cosmology in Scalar Tensor 
Gravity}, Fundamental Theories of Physics Series vol. 139 (Kluwer 
Academic, Dordrecht, 2004).

\bibitem{nohair1} S.W. Hawking, {\em Commun.\ Math.\ Phys.} {\bf 25}, 167 
(1972).

\bibitem{nohair2}T.P. Sotiriou and V. Faraoni, {\em Phys.\ Rev.\ Lett.} 
{\bf 108}, 081103 (2012).

\bibitem{nohair3}S. Bhattacharya, K.F. Dialektopoulos, A.E. Romano, and 
T.N. Tomaras, {\em Phys.\ Rev.\ Lett.} {\bf 115}, 181104 (2015).

\bibitem{nohair4}V. Faraoni, {\em Phys.\ Rev.\ D} {\bf 95}, 124013 
(2017).

\bibitem{JB} V. Faraoni, F. Hammad, A.M. Cardini and T. Gobeil, {\em 
Phys. Rev. D} {\bf 97}, 084033 (2018).

\bibitem{ETH1} D. Psaltis, arXiv:1806.09740 [astro-ph.HE].

\bibitem{ETH2} K. Akiyama {\it et al.} [Event Horizon Telescope 
Collaboration], {\em Astrophys. J.} {\bf 875}, L1 (2019).

\bibitem{gwtests1} J.R. Gair, M. Vallisneri, S.L. Larson, and J.G. Baker, 
{\em Living Rev. Relativity} {\bf 16}, 7 (2013).

\bibitem{gwtests2} J. Sakstein and B. Jain, {\em Phys. Rev. Lett.} {\bf 
119}, 251303 (2017).

\bibitem{gwtests3} P. Creminelli and F. Vernizzi, {\em Phys. Rev. Lett.} 
{\bf 119}, 251302 (2017).

\bibitem{gwtests4} T. Baker, E. Bellini, P.G. Ferreira, M. Lagos, J. 
Noller, and I. Sawicki, {\em Phys. Rev. Lett.} {\bf 119}, 251301 (2017).

\bibitem{gwtests5} T. Baker, E. Bellini, P.G. Ferreira, M. Lagos, J. 
Noller, and I. Sawicki, {\em Phys. Rev. Lett.} {\bf 119}, 251301 (2017).

\bibitem{gwtests6} L. Lombriser and A. Taylor, {\em J. Cosmol. Astropart. 
Phys.} {\bf 1603}, 31 (2016).

\bibitem{gwtests7} L. Lombriser and N.A. Lima, {\em Phys. Lett. B} {\bf 
765}, 382 (2017).

\bibitem{a} L. Iorio, {\em Universe} {\bf 1},38 (2015).
 
\bibitem{b} I. Debono and G.F. Smoot, {\em Universe} {\bf 2}, 23 (2016).

\bibitem{c} R.G. Vishwakarma, {\em Universe} {\bf 2}, 16 (2016).

\end{thebibliography}
\end{document}